# Negative-ion surface production in hydrogen plasmas: determination of the negative-ion energy and angle distribution function using mass spectrometry

J. P. J. Dubois[1], K. Achkasov[1,2], D. Kogut[1], A. Ahmad[1], J. M. Layet[1], A. Simonin[2], G. Cartry[1]

[1] *Aix-Marseille University, CNRS, PIIM, UMR 7345, 13013 Marseille, France*
[2] *CEA, IRFM, F-13108 Saint-Paul-lez-Durance, France*

## Abstract

This work focuses on the understanding of the production mechanism of negative-ions on surface in low pressure plasmas of $H_2/D_2$. The negative ions are produced on a HOPG sample (Highly Oriented Pyrolitic Graphite) negatively biased with respect to plasma potential. The negative ions created under the positive ion bombardment are accelerated towards the plasma, self-extracted and detected according to their energy and mass by a mass spectrometer placed in front of the sample. The shape of the measured Negative-Ion Energy Distribution Function (NIEDF) strongly differs from the NIEDF of the ions emitted by the sample because of the limited acceptance angle of the mass spectrometer. To get information on the production mechanisms, we propose a method to obtain the distribution functions in energy and angle (NIEADFs) of the negative-ions emitted by the sample. It is based on an *a priori* determination of the NIEADF and on an *a posteriori* validation of the choice by comparison of the modelled and experimental NIEDFs.

Corresponding author: gilles.cartry@univ-amu.fr

# 1. Introduction

Negative-ion (NI) production in low-pressure plasma has many applications and is studied in various areas, such as microelectronics[1,2,3,4,5,6], magnetically confined fusion[7,8], space propulsion[9,10,11] and sources for particle accelerators[12,13,14]. NI in low pressure plasmas are usually created by dissociative attachment of electrons on molecules (volume production)[15,16,17], but they can also be created on surfaces by the bombardment of positive ions or hyperthermal neutrals[18,19,20,21,22,23,24]. Volume-produced NI sources are mainly used in microelectronic [1,2,3,4,5,6] and space propulsion applications[9,10,11] while surface production of NI is employed in magnetically confined fusion devices[7,8] and sources for particle accelerators[12,13,14]. Surface production of negative-ions is also observed in sputtering processes as shown in 25,26,27.

In magnetically-confined fusion devices (tokamaks), NI can be used to generate a fast neutral beam through interaction with a stripping gas target; such beam is used to heat the plasma and to implement the current drive. Given the huge dimensions of the ITER device and its successor DEMO, neutral beam energies in the range of 1–2 MeV are required; at these energies the neutralization of positive ions becomes very inefficient. Indeed, the yield of neutralization of the positive ions by a $D_2$ gas tends to zero above 100 keV, while its value is around 55% at 1 MeV for NI[28]. Therefore, there is a great research effort dedicated to the development of a high current NI source (50A $D^-$ beam for ITER)[29,30,31,32,33,34,35,36,37]. This source uses cesium injection inside the plasma in order to increase strongly the NI extracted current[38,39,40]. However, as the use of cesium complicates noticeably the neutral beam injection device, there is a demand for the development of cesium-free NI sources in $H_2/D_2$ plasmas[41] or for the reduction of the cesium consumption in fusion negative-ion sources[42]. Several papers dedicated to this subject have been recently published[19,42,43,44, 45,46,47,48,49]. In the present paper, we are pursuing our studies on NI surface production in cesium-free low-pressure $H_2/D_2$ plasmas.

In our experimental set-up[47] a sample is introduced in the plasma and negatively biased with respect to the plasma potential. The negative ions formed on the surface are accelerated by the sheath in front of the sample and self-extracted on the other side of the plasma towards a mass spectrometer. Negative ions are detected according to their energy and mass and Negative-Ion Energy Distribution Function (NIEDF) is measured. HOPG (Highly Oriented Pyrolitic Graphite) has been chosen as a reference material for the study, because of its high yield of production of negative-ions and because of its simplicity to cleave and clean. In order to gain an insight on the mechanisms of the NI surface production, it is important to analyze and characterize the shapes of the measured NIEDFs.

It has been shown that the measured NIEDFs strongly differ from those of the ions emitted by the sample surface because of the limited acceptance angle of the mass spectrometer[47]. Furthermore, the measurements provide only information about energy and mass of the negative ions, and no information is obtained concerning their angle of emission from the surface. To get information on the production mechanisms, it is mandatory to determine the distribution functions in energy and angle of the negative-ions (NIEADFs) emitted by the sample. A model has been developed to obtain the NIEADFs earlier[47]. The principle was to choose *a priori* the NIEADF $f(E,\theta)$ and to compute NIEDF from the chosen NIEADF. The good agreement between the NIEDFs measured by the mass spectrometer and the modelled ones allowed us to validate the choice of the chosen NIEADF. However, it was shown that only few ions among all the emitted ones are collected by the mass spectrometer (5% of collection), hence a good agreement between the model and the experiment only validated the choice of a small part of the NIEADF. Indeed, because of the low acceptance angle of the mass spectrometer, only ions emitted at small angles with respect to the axis of the mass spectrometer can be collected. In the present paper NIEDFs are measured for different tilt angles of the sample, allowing the collection of ions emitted from the surface at any angle and energy. The corresponding NIEDFs are then computed using the chosen NIEADF and compared to the experimental ones in order to further validate the choice of the NIEADF.

The paper is organized in three parts. In the first one, the experimental set-up is briefly described. The second part is dedicated to the improvement of the model developed previously[47], which allows taking into account a tilt of the sample with respect to the mass spectrometer axis. In the last part the computed NIEDFs are presented and compared to the experiments. Finally, the other outputs of the model are presented and discussed.

## 2. Experimental set-up

The reactor, diagnostic instruments and plasma conditions used were described in detail elsewhere[47]. The measurements are performed in a helicon reactor consisting of an upper cylindrical source chamber (Pyrex tube, 360 mm long and 150 mm diameter) and a lower spherical diffusion chamber (stainless steel chamber, radius 100 mm). The plasma is created in the source chamber by an external antenna surrounding the Pyrex tube and connected to a 13.56 MHz RF generator. The HOPG sample is inserted in the diffusion chamber. Its surface exposed to the plasma is a disc of 8 mm in diameter facing the mass spectrometer nozzle located at 37 mm away. The mass spectrometer axis passes through the center of the sample and the latter can be rotated in the direction perpendicular to this axis (Figure 1). It has been found that the rotation of the sample strongly reduces the intensity of the recorded NIEDFs. Increasing the RF power applied to the plasma to increase both the positive-ion flux and the resulting NI flux is not a satisfying solution since it also increases the RF oscillations of the plasma potential, perturbing the measurements. Therefore, instead of the RF source used in a capacitive mode to generate the plasma, an ECR source (Electron Cyclotron Resonance) from Boreal Plasma was installed in the diffusion chamber, 5 cm away from the sample. The ECR source was operated at 2.45 GHz, which gives electron resonance for a B field of 845 Gauss (such value is reached less than one centimeter away from the magnet). A more detailed description of the ECR source can be found in [16]. The plasma density was increased by one order of magnitude and the RF fluctuations problem was avoided. It has been checked that NIEDF shapes are not affected by the magnetic field of the ECR source, which is around 50 gauss at the sample surface. In addition, the orifice diameter in the mass spectrometer nozzle has been increased from 35 μm to 100 μm. This leads to an increase of the signal by a factor ~10. Moreover, the clamping part of the sample holder (see Figure 1 in [47]) has been made much thinner (0.1 mm) improving measurements at high tilt angle when the signal is low.

Measurements were performed at 1 Pa $H_2$ and 60 W of injected power. The plasma density, as measured by Langmuir probe, is $2.5 \cdot 10^{15}$ m$^{-3}$, the electron temperature is 1 eV and the plasma potential is 7 V. The positive ion flux composition is determined by mass spectrometry. $H_3^+$ flux represents around 75 % of the total ion flux, $H_2^+$ flux around 14% and $H^+$ 11%. As explained in 19, due to the variation of the transmission probability of the mass spectrometer, this is only an estimation of the ion fractions, which can vary by few percent depending on the mass spectrometer tuning. The sample was biased to $V_s$ = -130 V leading to the dominant positive ion impact energy of 45 eV per proton. The energy of H$^-$ ions entering the plasma is between 130 eV (ion created at rest on the surface) and 175 eV (ion created with initial energy of 45 eV, see Figure 2). In this energy range the electron detachment cross section (H$^-$ + $H_2$) is between $5.3 \cdot 10^{-20}$ and $5.9 \cdot 10^{-20}$ m$^2$ [50], giving a mean free path for negative-ion detachment in the range of 70–80 mm. As the distance between the sample and the mass spectrometer is 37 mm, this leads to a loss of around 40% of the ion flux. However, as the detachment cross section is largely dominating over other processes (such as momentum transfer[50]), and because detachment cross section is not varying too much (from $5.3 \cdot 10^{-20}$ to $5.9 \cdot 10^{-20}$ m$^2$), this does not affect the NIEDF shape as demonstrated previously[46].

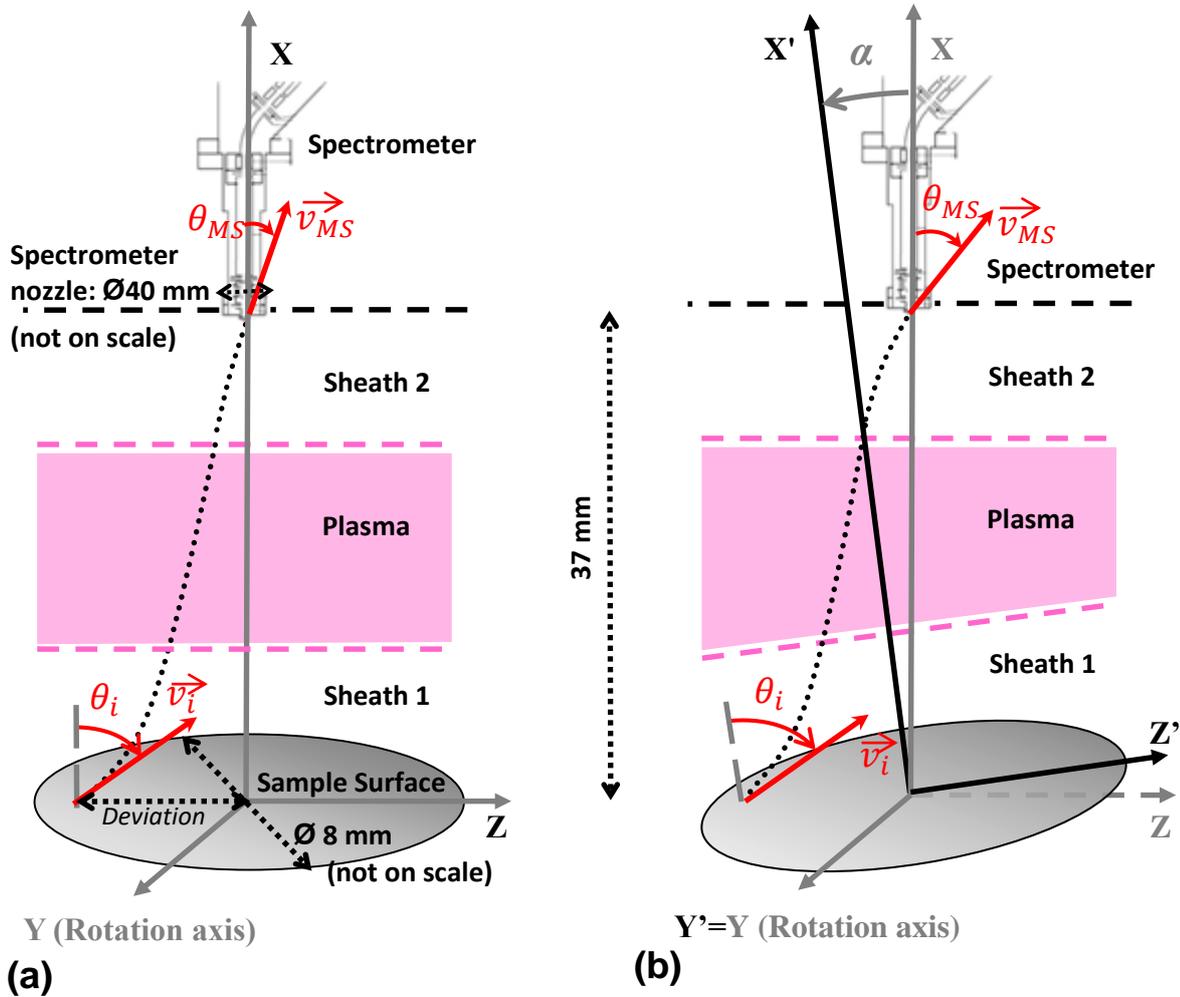

Figure 1. Sketches of the NI trajectory between the sample surface and the mass spectrometer for the tilt angle $\alpha$ = 0 (a) and $\alpha \neq 0$ (b). The figure is not on scale, the sample size is enlarged.

## 3. Modelling

As explained in the introduction, a model has been previously developed to obtain the NIEADF on the sample from the NIEDF measured by the mass spectrometer. The principle is to choose *a priori* the NIEADFs $f(E,\theta)$ and to validate *a posteriori* this choice. In this model the NI trajectories between the sample and the mass spectrometer are computed based on their energy and angle of emission. All ions missing the entrance of the mass spectrometer or arriving to it with an angle $\theta_{MS}$ higher than the acceptance angle $\theta_{aa}$ are eliminated from the calculations (Figure 1). The energy and angle distribution of the remaining ions is labelled $f'(E,\theta)$. The transmission probability of the ions passing through the mass spectrometer itself is calculated using the SIMION software[51]. The energy distribution function of the ions collected by the mass spectrometer $f''(E)$ is obtained by summing up all the ions over the angles weighted by their transmission probability at a given angle and energy. This NIEDF is then compared with the experimental one to validate the choice of $f(E,\theta)$, and if required the method is iterated. In our previous paper[47] we have shown that the distribution $f'(E)$ at the mass spectrometer nozzle is not strongly different from the distribution at the mass spectrometer detector $f''(E)$. Therefore, in order to speed up the calculations, only $f'(E)$ is considered in the present paper.

In previous papers we have shown that the negative-ions are formed by the backscattering of positive ions as NI and by the sputtering of adsorbed hydrogen (deuterium) atoms as NI[19,43,44]. The energy and angular distribution of backscattered and sputtered particles computed by the SRIM code[52] has been chosen as the initial NIEADF $f(E,\theta)$. As SRIM does not take into account the surface ionization, this choice implicitly assumes that the surface ionization probability is independent of the angle and energy of the particles emitted. On metals, the ionization probability is usually dependent on the outgoing velocity[53,54,55,56,57]. Assuming a constant ionization probability is a strong assumption that has been largely discussed previously[47]. Positive ions impacting the surface have been divided in the modelling in three groups corresponding to the three ion types observed experimentally: $H_3^+$, $H_2^+$, $H^+$. To take into account molecular ion dissociation at impact, the $H_3^+$ ion is modelled by proton impact at one third of $H_3^+$ energy with a flux three times higher than the $H_3^+$ one (3 protons at 45 eV for each $H_3^+$ impact). In the same way, $H_2^+$ is modelled by the impact of protons at half the $H_2^+$ energy and twice the $H_2^+$ flux (2 protons at 67 eV for each $H_2^+$ impact). The sample material is assumed to be an amorphous *a*-C:H layer (30% H), since the graphite surface exposed to plasma is subjected to hydrogen implantation and defect creation in the subsurface layer, which has been confirmed by Raman spectroscopy measurements[45,46]. The parameters of the SRIM calculation are listed elsewhere[48] and reminded in Table 1. $10^7$ incident positive ions have been used in these calculations yielding in ~$2\cdot10^6$ H particles emitted from the surface, including backscattered and sputtered ones. The acceptance angle $\theta_{aa}$ as computed by SIMION is 1° with the present mass spectrometer settings.

| Species | % in the layer | Displacement energy (eV) | Lattice energy (eV) | Surface binding energy (eV) |
|---|---|---|---|---|
| C | 70 | 25 | 3 | 4.5 |
| H | 30 | 2.5 | 3 | 3 |

Table 1: Parameters used for the SRIM calculations. *a*-C:H layer is modelled with a density of 2.2 g/cm$^3$. The type of calculation used is "Surface sputtering / Monolayer Collision Steps". $10^7$ proton impacts with incidence angle of 0° (with respect to the sample normal) are simulated. The impact energies are given in the text.

Figure 2 shows a comparison between the measured (black) and the computed (red for the case of 100% $H_3^+$ or blue for the case of 75% $H_3^+$, 14% $H_2^+$, 11% $H^+$) NIEDFs in the ECR plasma. When the three populations of positive ions are taken into account, the model reproduces quite well the shape of the NIEDF, which presents a main peak at low-energy (0–10 eV), a tail with a slight decreasing slope at intermediate energy (10–30 eV) and a slope breaking around 30 eV followed by the high energy tail. In case when only the $H_3^+$ ion flux, which is dominating under the present experimental conditions, is considered, the measured distribution is globally well fitted by the model with the exception of the high-energy tail (above 30 eV). It is due to the fact that only $H_2^+$ and $H^+$ ions with higher energy per proton can contribute to the formation of negative-ions with such high energy (see considerations on the maximum H$^-$ energy in [19]). Let us note however that the modelling with $H_3^+$ positive ion only gives a good agreement for 95% of the negative-ion population (only few negative-ions have energy higher than 35 eV). Finally, Figure 2 demonstrates that only a part of the emitted ions is collected by the mass-spectrometer: the distribution function of the collected ions *f'(E)*, shown with empty symbols, differs strongly from the distribution function of the emitted ions *f(E)*, shown with filled symbols.

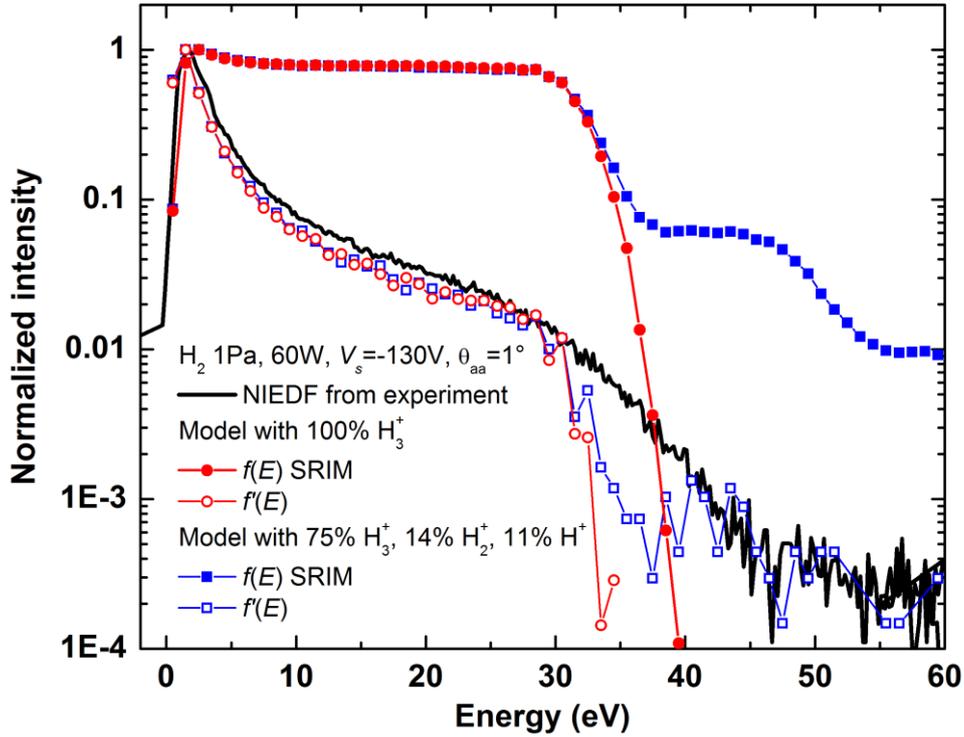

Figure 2. Comparison between the calculated energy distribution function $f'(E)$ of the negative ions collected by the mass spectrometer (empty symbols) and the experimental one (black line) obtained at 1 Pa, 60 W with the ECR source. Red circle correspond to the case of 100% $H_3^+$, blue squares to the case of 75% $H_3^+$, 14% $H_2^+$ and 11% $H^+$. The energy distribution functions of ions on the sample $f(E)$ calculated by SRIM and used as input in the model are shown with filled symbols. All NIEDFs are normalized to the peak value.

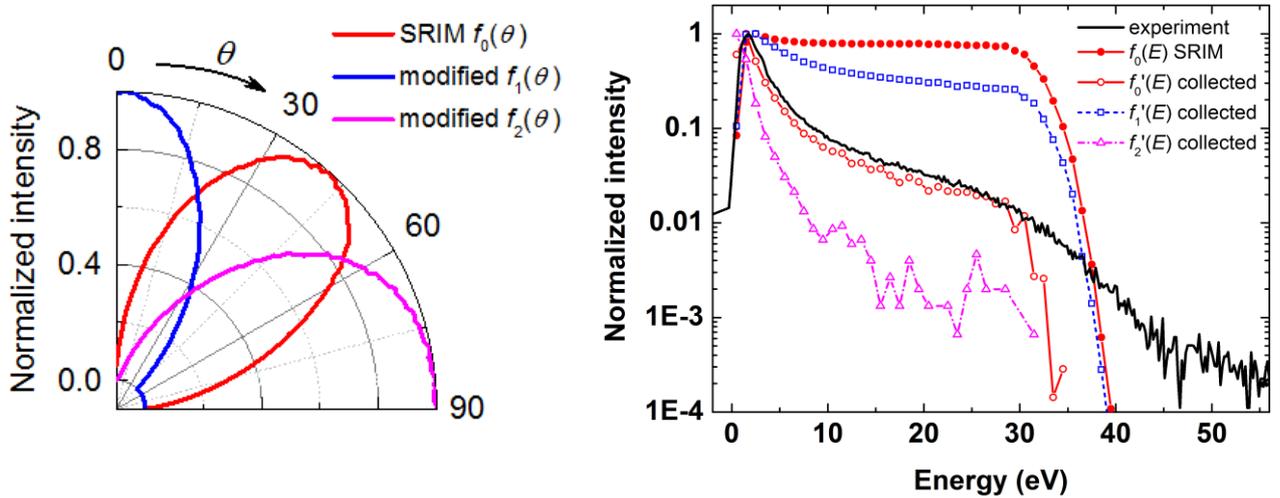

Figure 3. Left: normalized angular distributions $f(\theta)$ of ions leaving the sample surface, either given by SRIM (red) or artificial (blue and magenta). Right: corresponding energy distribution functions of the collected NI $f'(E)$ given by the model (empty symbols) compared to the experimental curve (black). The input of the model is the same $f(E)$ for all cases, as given by SRIM when only $H_3^+$ is considered (filled red symbols). All NIEDFs are normalized to the peak value.

Left: normalized angular distributions $f(\theta)$ of ions leaving the sample surface, either given by SRIM (red) or artificial (blue and magenta). Right: corresponding energy distribution functions of the collected NI $f'(E)$ given by the model (empty symbols) compared to the experimental curve (black). The input of the model is the same *f(E)* for all cases, as given by SRIM when only $H_3^+$ is considered (filled red symbols). All NIEDFs are normalized to the peak value.Figure 3 demonstrates the sensitivity of the modelled NIEDF *f'(E)* to the input angular distribution *f(θ)* of ions leaving the sample surface. As it will be shown in the following section, only the ions with polar angle of emission *θ* close to zero can be collected when the sample is perpendicular to the mass spectrometer axis. Artificial shift of *f(θ)* given by SRIM towards *θ* = 0 favors collection of NI in the whole range of energies (see blue curves on the left and on the right), while shifting *f(θ)* to 90° leads to a considerable decrease of collection at higher energies (magenta curves). Therefore, the shape of the resulting *f'(E)* is highly sensitive to the assumed *f(E,θ)* at the sample surface and can be used to validate *a priori* the choice of *f(E,θ)*.

The sheaths in front of the sample and in front of the mass spectrometer can be considered planar in our experimental conditions[47] so that the electric field vector is reduced to one component in the X direction (Figure 1). Indeed, the sheath thickness is 4.5 mm while the 8 mm diameter sample is placed at the centre of a 20 mm × 35 mm surface defined by the clamp size. As this surface is much larger than the sheath size we can assume that sheath edge effects concern only the clamp edges and not the sample. Furthermore, as the clamp has been made very thin (0.1 mm), we can consider that the sheath is planar above the whole surface of the sample. Same reasoning shows that the sheath in front of the mass spectrometer nozzle can also be considered as planar. The potential variation along the X axis in the sheaths is calculated from the Child Langmuir law. Then a simple calculation of the ion trajectories in the sheath can be done. The input parameters for the trajectory calculations, such as the electron density, the electron temperature, the plasma potential and the applied surface bias which allow the computation of the sheath potential, are taken from the experiment (see section 2, "Experimental set-up"). These parameters correspond to the ECR $H_2$ plasma at 1 Pa and 60 W. These calculations have already been detailed in the previous paper[47] for the sample perpendicular to the mass spectrometer axis. The same method is used here for a tilted sample. Each NI ejected from the surface is characterized by its kinetic energy *E* and the angles of emission: polar angle *θ* and azimuth angle *φ* in spherical coordinate system associated with the sample surface. The distribution in *E, θ, φ* is taken from the SRIM calculations; note that the negative ions are emitted with a uniform distribution in *φ*. *α* is the tilt angle between the normal of the sample and the mass spectrometer axis. For *α* = 0° there is a cylindrical symmetry around the mass spectrometer axis, and the angle *φ* has no importance in the calculations. For *α* ≠ 0° there is no more symmetry and the angle *φ* matters. Two output values are mandatory to determine if an ion can be collected by the mass spectrometer:

i) the arrival angle $θ_{MS}$ of the negative ion at the mass spectrometer entrance. This angle is compared to the mass spectrometer acceptance angle $θ_{aa}$ determined by the SIMION simulations. The ion is eliminated from the simulation if $θ_{MS}$ is higher than $θ_{aa}$.

ii) the lateral deviation, which is the distance in the YZ plane (plane perpendicular to the mass spectrometer axis, see Figure 1) between the starting location of the negative ion on the sample and its arrival location in the plane of the mass spectrometer. When the sample is not tilted, the lateral deviation must be lower than the sample radius (4 mm) otherwise the ion is eliminated from the calculation. Indeed, an ion emitted from the sample surface cannot reach the mass spectrometer entrance hole if it is laterally deviated by more than 4 mm. When the sample is tilted, the deviation must verify the equation $(\frac{Dev_y}{R})^2 + (\frac{Dev_z}{R \cos\alpha})^2 \leq 1$ where $Dev_y$ and $Dev_z$ are the deviation components in the YZ plane, $R$ is the sample radius. This relation describes an ellipse which is the projection of the sample surface (disc) in the plane perpendicular to the mass spectrometer nozzle.

## 4. Results and discussion

As it is mentioned above, only the negative ions with certain energies and angles of emission can be detected by the mass-spectrometer; the accepted energies and angles vary with the tilt of the sample $\alpha$. This is illustrated by Figure 4 which shows polar plots of emitted NI (grey dots) and collected NI (blue dots) for different $\alpha$. The radial position gives the NI emission energy normalized to the impact energy $E_{\text{impact}} = 45$ eV of the impinging positive ions (only $H_3^+$ ions are considered in this calculation). The angular position gives the emission angle of a NI with respect to the normal of the sample. Angle and energy of emission are obtained from the SRIM calculations (emitted ions) and from the modelling (collected ions). The normalized angular distribution of either emitted NI (black line) or collected NI (red line) is also shown in Figure 4.

The maximum of the emitted NI distribution is always at $\theta = 45°$, however the maximum of the collected NI distribution depends on $\alpha$; for example, at $\alpha = 0°$ it is located at $\theta = 4°$, although the range of collection extends up to $\theta = 27°$. It may be explained by the fact that the ions emitted from the surface with small angles and energies have trajectories which are easily rectified by the electric field present in the sheaths; hence, NI arrive to the mass spectrometer entrance with the angle $\theta_{MS} < \theta_{aa}$ and are collected. The trajectories of the ions emitted at high energies are not rectified efficiently unless $\theta$ is close to $0°$. Hence in the experiment one favors the collection of NI emitted with small angles $\theta$ and/or small energies at $\alpha = 0°$.

When the sample is tilted, the collected ions come from a different part of the initial distribution $f(E,\theta)$. This may be observed in Figure 4 where the polar plots for $\alpha = 5°$, $10°$, $15°$ show the shift of the maximum of the angular distribution of the collected NI to higher $\theta$. Therefore, by tilting the sample it is possible to collect the NI of all emission angles and energies. It is important to note that the whole distribution of emitted ions is scanned (not shown here) when using $\alpha$ angles from $0°$ to $30°$ in the case of proton impact at 45 eV ($H_3^+$ impacting on the surface at 135 eV).

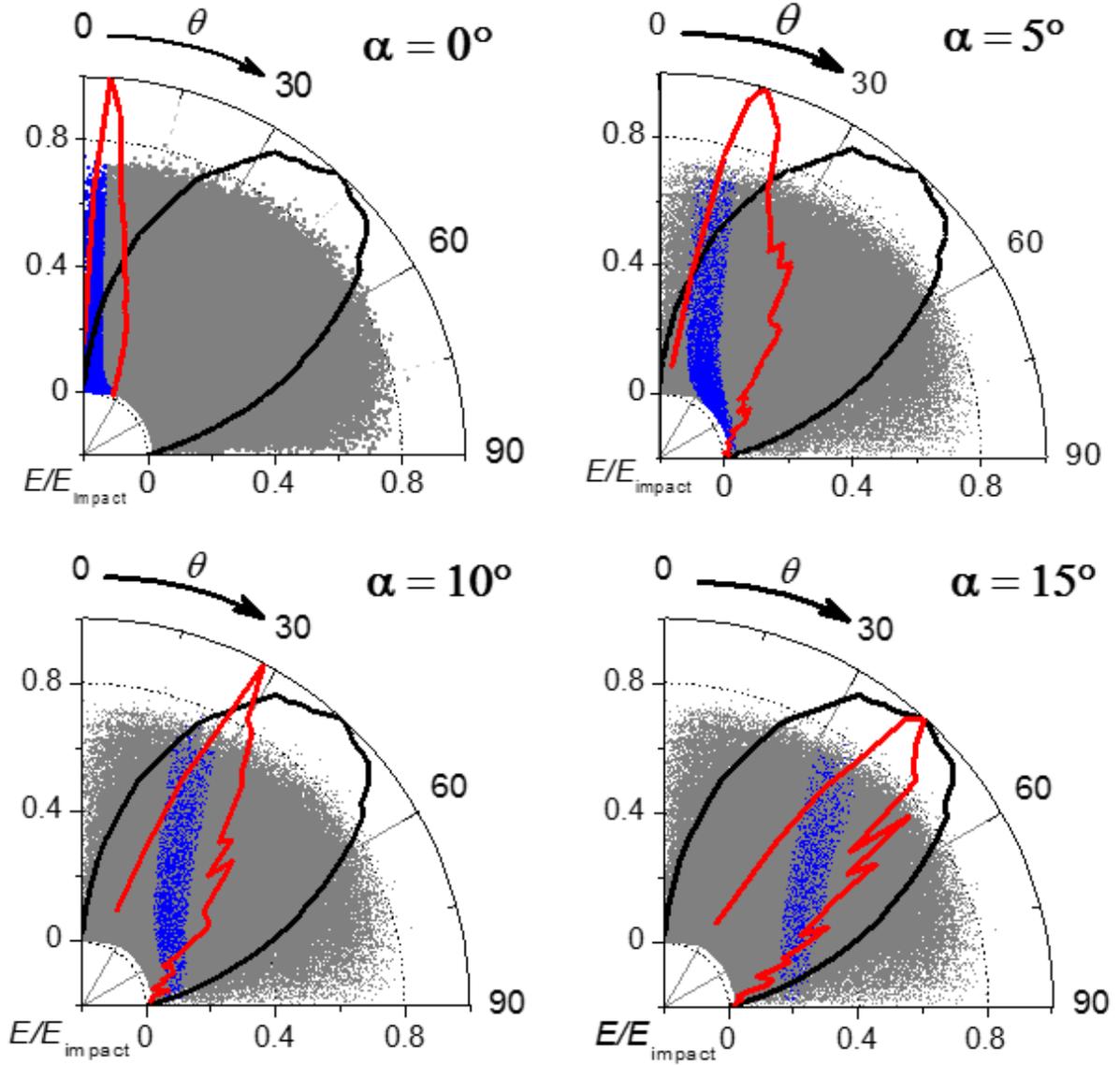

Figure 4. The polar plots of NI emitted from HOPG surface ($f(E,\theta)$, grey dots, from SRIM calculations) and NI collected by the mass-spectrometer ($f'(E,\theta)$, blue dots) as given by the model for different angles $\alpha$ of the sample tilt. The surface bias is $V_s = -130$ V. The radial position gives the negative-ion emission energy normalized to the impact energy of the positive ions $E_{impact} = 45$ eV. Black line is the normalized angular distribution of the emitted negative ions. Red line is the normalized angular distribution of the collected negative ions. Input parameters of the modelling correspond to the situation of an ECR $H_2$ plasma at 1 Pa, 60 W, Vs = -130 V.

The NIEDFs from the experiment and the model are compared for different tilt angles of the sample from 0° to 20° in Figure 5. The peak of the NIEDF measured at $\alpha = 0°$ is normalized to unity, and the NIEDFs measured at other tilt angles are normalized accordingly. The peak of the distribution computed for $\alpha = 2°$ is normalized to the experimental one measured at $\alpha = 2°$ and the other computed distribution are normalized accordingly. This choice (normalization at $\alpha = 2°$ rather than at $\alpha = 0°$) is discussed below.

The shapes of the distributions at any alpha angle are well reproduced by the model. When $\alpha$ increases, the energy onset of the distribution shifts to higher energies. These shifts in the measured NIEDFs are reproduced by the model. The global experimental intensities are decreasing with $\alpha$ and this decrease is also well reproduced by the model with one exception at $\alpha = 0°$ (see the discussion below). The distributions measured at different alpha always superimpose with the tails of the previous ones. This is also reproduced by the model. The agreement between the experiment and the calculation is good enough to conclude that the improved model can be used to simulate the trajectories with a tilted sample. It validates the use of the distribution function $f(E,\theta)$ calculated by SRIM for all the values of energy and angle of the emitted NI. Up to now this choice was validated for small angles only since at $\alpha = 0°$ only ions emitted at small angles are collected[47] (Figure 4). Thus, for HOPG, the surface-produced NI distribution can be approximated by the distribution of neutrals calculated by SRIM with the parameters listed above.

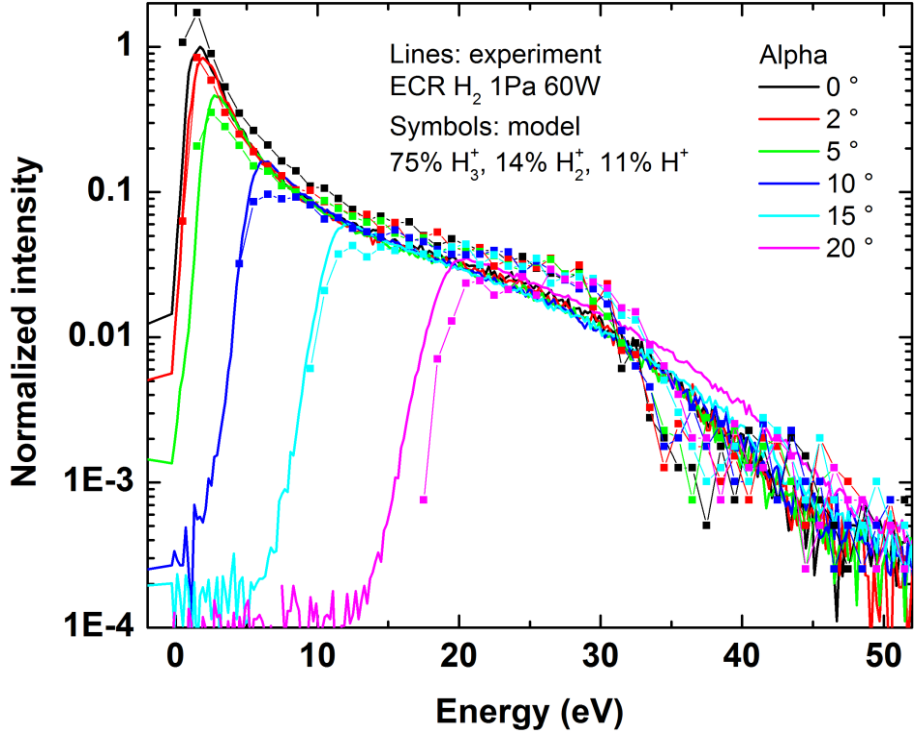

Figure 5. Comparison of the experimental NIEDF (solid lines) with the modelled ones (symbols) for different tilts of the sample (from $\alpha = 0°$ to 20°, with a step of 5°). ECR $H_2$ plasma 1 Pa, 60 W, $V_s = -130$ V.

One can notice that the variations in the distribution intensity with $\alpha$ is well reproduced by the model except for $\alpha = 0°$. For $\alpha = 0°$, the intensity calculated by the model is higher than the experimental one. In other words, the drop in intensity between $\alpha = 0°$ (symmetrical situation) and $\alpha \neq 0°$ (non-symmetrical situation) predicted by the modelling is observed in the experiment but is less important than expected. We have no explanation at the moment. It could arise from imperfect determination of the angle $\alpha = 0°$ in the experiment. Further studies are required to clarify this point. Let us note that the normalization between model and experiment for $\alpha = 0°$ (as done on Figure 2) would have given modelled NIEDFs with lower intensity than the experimental ones for all angles $\alpha > 0°$.

The decrease of the NIEDF peak intensity with the growth of $\alpha$ can be understood by simple geometrical considerations. For $\alpha = 0°$ there is a cylindrical symmetry around the mass spectrometer axis. The criteria of collection of an ion by the mass spectrometer only depends on $E$ and $\theta$, there is no restriction on $\varphi$. The ions collected for a given $(E,\theta)$ originate from a circle on the sample, i.e. $\varphi \in [0:2\pi]$. When the symmetry is broken ($\alpha \neq 0°$), only a part of the distribution in $\varphi$ is collected, so the signal strongly decreases. In order to illustrate this, the following calculation was performed. A distribution of ions emitted from the surface was created, each ion having a unique combination of $E$, $\theta$ and $\varphi$ with $E$, $\theta$ and $\varphi$ distributed uniformly (from 0 to 50 eV, 0 to 90°, 0 to 180° correspondingly). Transmission of these ions in the plasma was calculated by the model for different $\alpha$. As a result, one could construct the distribution $f'(\theta, \varphi)$ for each energy individually from calculations for different sample tilts $\alpha$. Two examples of $f'(\theta, \varphi)$ are shown in Figure 6: for $E = 2$ eV on the left and for 17 eV on the right. Each dot in Figure 6 corresponds to an ion and its color indicates the tilt angle $\alpha$ at which this ion can be measured. One can see that the measured range of $\varphi$ is strongly reduced for higher energies and/or higher $\alpha$ and $\theta$. One can also note that the minimum number of different sample tilts $\alpha$, which has to be measured to cover the total range of $\theta$ at a given energy, increases with energy.

The fact that at higher values of $\alpha$ the angular distribution of the collected NI shifts to higher $\theta$ (Figure 4, Figure 6) has an important consequence. It has been shown before that NI originate from both sputtered and backscattered particles from the sample surface induced by the positive ion bombardment and the angular emission yield is strongly dependent on the process considered[48]. The emission yield for sputtered particles shows a maximum at $\theta = 20°$ and goes to zero above 45°; as far as backscattering is considered, the yield is maximal at 45° and goes to zero above 70° [48]. Therefore, by varying the tilt angle $\alpha$ we can collect preferentially either sputtered or backscattered NI; this is illustrated in Figure 7.

Figure 8 shows the spatial distribution of the original locations of NI on the sample surface collected for each tilt angle. The spatial origin of the ion is simply inferred from its deviation which is computed based on $E$, $\theta$, $\varphi$ of the particle. A uniform distribution of $E$, $\theta$, $\varphi$ has been chosen for this calculation as an example of an arbitrary NIEADF, in order to enlarge the discussion from the NI surface production on HOPG to the general case.

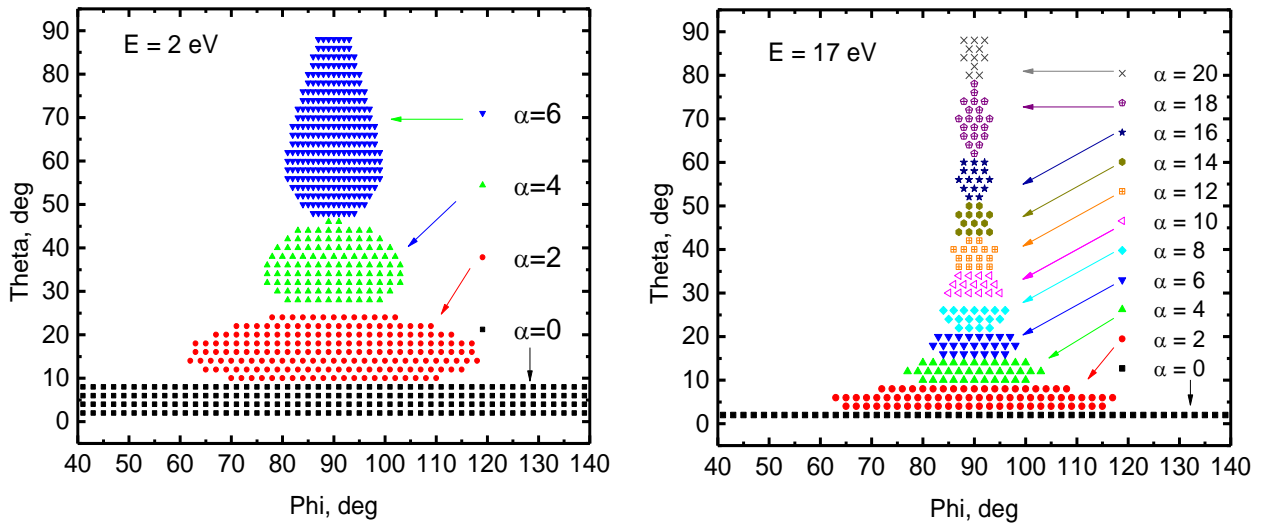

Figure 6. Distribution $f'(\theta, \varphi)$ of the NI collected by the MS obtained from different sample tilts $\alpha$ for a given energy: $E = 2$ eV (left) and $E = 17$ eV (right). The distribution of emitted ions has been chosen uniform in $E$, $\theta$, $\varphi$ for this calculation. Input parameters of the modelling correspond to an ECR $H_2$ plasma at 1 Pa, 60 W, $V_s = -130$ V.

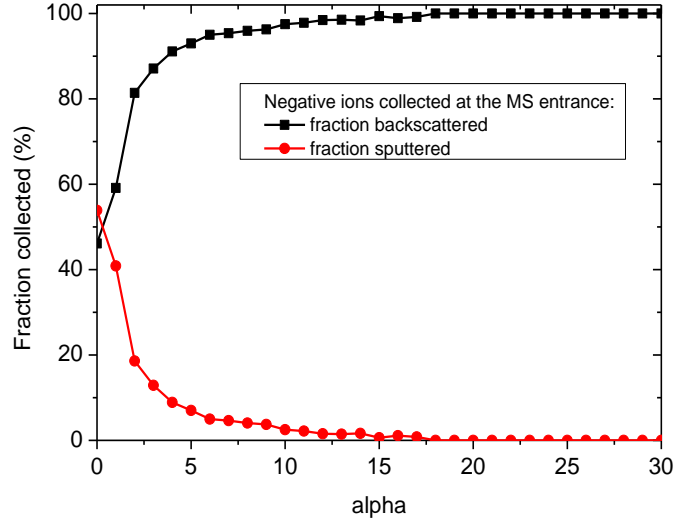

Figure 7. Fraction of sputtered and backscattered NI, which are collected by the mass-spectrometer, for different tilts of the sample $\alpha$; the calculation is done in SRIM with $E_{impact}$ = 45 eV. Input parameters of the modelling correspond to the situation of an ECR $H_2$ plasma at 1 Pa, 60 W, $V_s$ = -130 V.

The ions collected at $\alpha = 0°$ originate from a circle on the sample (see Figure 8). When the symmetry is broken ($\alpha \neq 0°$), only a part of the distribution in $\varphi$ is collected and the signal strongly decreases. Moreover, the spot of origin of the ions transforms into an ellipse and shifts in the direction of the mass spectrometer, perpendicular to the rotation axis, see Figure 8. The spot remains small in dimensions (~2 mm) compared to the total sample surface (8 mm in diameter). Despite the shift, it can be seen that even at the tilt angle $\alpha = 30°$, NI never originate from the sample holder (or even the edge of the sample), which justifies our measurement scheme. Figure 8 shows that it would be possible to work with smaller samples if required.

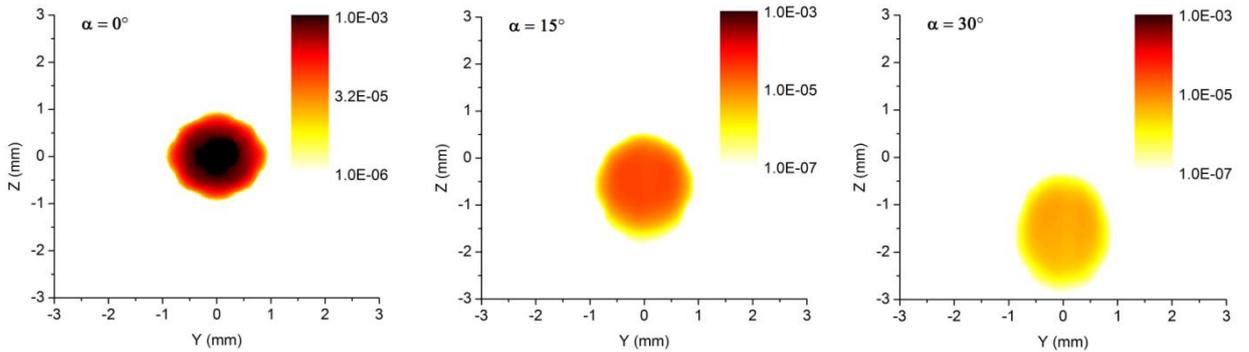

Figure 8. Maps showing the spatial distributions of the origin on the sample surface of NI collected by the mass spectrometer. The maps are drawn for tilts of the sample $\alpha = 0°$, 15° and 30°. Uniform distribution in $E$, $\theta$, $\varphi$ is assumed. Input parameters of the modelling correspond to an ECR $H_2$ plasma at 1 Pa, 60 W, $V_s$ = -130 V.

The main output of the present work is the experimental confirmation that backscattered and sputtered neutral distribution functions as computed by SRIM can be used for negative-ions. It implies that the surface ionization probability under the present experimental conditions is not dependent on the angle and energy of the outgoing particle. This is not usually the case for metals, while it can be possible for insulators. Further studies are required to clarify the reason for this independence. In any case, knowing the NIEADF on the sample surface is of primary importance for the study of negative-ion surface production in plasmas with the aim of finding proper negative-ion enhancer materials. Figure 2 shows that at any tilt angle the fraction of ions collected is low compared to the total number of emitted ions. Measuring NIEDF at one particular tilt angle is therefore not enough to test the efficiency of one particular material with respect to the negative ion surface production. Hydrogenated materials, such as carbon considered here, will probably present a high signal at 0° tilt angle due to the preferential collection of sputtered ions (Figure 7) despite most of the ions are created on the surface by backscattering[48]. Non-hydrogenated materials might present a small signal at 0° tilt angle despite a high efficiency. Comparison of two different materials in terms of negative-ion yield must therefore be done at all tilt angles. The best approach is to use these measurements and the model to validate NIEADF given by SRIM and to compare directly NIEADF of both materials. Let us finally note that the ions formed at low energy and low angle are preferable in a negative-ion source in order to obtain well collimated beams of extracted ions. Thus, sputtering process is preferable since it gives lower energy and lower average angle emitted ions. In that sense, hydrogenated materials are interesting for the NI surface production.

## 5. Conclusion

The method developed here to determine the initial energy and angular distribution function (NIEADF) of NI emitted from the sample surface in $H_2$ plasma is validated by a good agreement of the model with the experiment. Indeed, using SRIM distribution function as initial NIEADF, the model can reproduce the Negative-Ion Energy Distribution Functions (NIEDFs) measured by the mass spectrometer at different tilt angles of the sample. Moreover, we showed by tilting the sample that this validation concerns the whole distribution of emitted ions in terms of energy and angle. It confirms that the NIEADF for HOPG is close to the neutral distribution function given by SRIM. It implies that the NI formation probability on the graphite surface, under the present experimental conditions, is almost independent of the angle and energy of emission as stated previously[47]. The present paper shows that only a small part of the negative-ions emitted by the sample is collected by the mass spectrometer and one measurement at one particular tilt angle is not enough to characterize negative-ion yield. Furthermore, the model shows that hydrogenated materials will probably present high negative ion signal at 0° tilt angle contrary to non-hydrogenated materials. Therefore, in order to analyze efficiency of different negative-ion enhancer materials, it is of primary importance to compare the corresponding NI yields at different tilt angles. The best approach is to use SRIM to compute NIEADF on the sample surface and to validate it by comparison with experimental NIEDF measured at different tilt angles. In that way, the NIEADF of ions emitted by different materials can be directly compared.

The present method for the investigation of surface production of negative-ions in plasma is not restricted to hydrogen ions and to the application to Neutral Beam Injectors for fusion. For instance, it can also be used for analysis of oxygen negative-ions formed during thin layer deposition by reactive sputtering [25,26,27].

# Acknowledgments

This work was carried out within the framework of the French Research Federation for Fusion Studies (FR-FCM) and of the EUROfusion Consortium. It has received funding from the Euratom research and training programme 2014-2018 under grant agreement No 633053. The views and opinions expressed herein do not necessarily reflect those of the European Commission. Financial support was received from the French Research Agency within the framework of the project ANR BLANC 13-BS09-0017 H INDEX TRIPLED.